\documentclass[preprint,aps,prd,showpacs,eqsecnum,nofootinbib]{revtex4}
\usepackage{graphicx}
\usepackage{rotating}
\usepackage{t1enc}

\def\gtap{\mathrel{ \rlap{\raise 0.511ex \hbox{$>$}}{\lower 0.511ex
   \hbox{$\sim$}}}} 
\def\ltap{\mathrel{ \rlap{\raise 0.511ex
   \hbox{$<$}}{\lower 0.511ex \hbox{$\sim$}}}}

\begin{document}

\title{An Appearance-Like Reactor Experiment To Measure $U_{e3}$}  
 
\author{
\mbox{Jos\'e Bernab\'eu$^{1}$} and
\mbox{Sergio Palomares-Ruiz$^{1,2,3}$}}

\affiliation{
\mbox{$^1$  Departamento de F\'{\i}sica Te\'orica and IFIC,
Universidad de Valencia-CSIC,}
\\
\mbox{46100 Burjassot, Valencia, Spain}
\\
\mbox{$^2$  Department of Physics and Astronomy, UCLA, Los Angeles, CA
  90095, USA}
\\
\mbox{$^3$  Department of Physics and Astronomy, Vanderbilt
  University, Nashville TN 37235, USA}}



\vspace{6mm}
\renewcommand{\thefootnote}{\arabic{footnote}}
\setcounter{footnote}{0}

\begin{abstract}
Conventional reactor neutrino experiments are dissapearance
experiments, and thus have less sensitivity to small mixing angles
than appearance experiments do. It has been recently shown that
future reactor neutrino experiments consisiting of a near and far
detector are competitive with first-generation superbeams in order to
determine $\sin^{2}{2\theta_{13}}$ down to $10^{-2}$. We show
that by using the antineutrino-electron elastic scattering at the near
detector around the configuration where
$d\sigma^{\overline{\nu}_e}/dT$ presents a dynamical zero, an
appearance-like experiment can be simulated, with a sensitivity
comparable to the one achieved with the inverse $\beta$-decay reaction
in the far detector. Thus, the near detector could also be used to
look for oscillations. We present how antineutrino-electron elastic
scattering could be properly used for this purpose allowing that the
combination of the measurements in the far detector and in the near
detector would push the sensitivity of the experiment to a lower value
of $\theta_{13}$.  
\end{abstract}

\pacs{14.60.Pq, 13.15.+g, 28.41.-i                 
\hspace{3cm} FTUV-03-1111 , IFIC-03-51 }

\maketitle


\section{Introduction}
The old solar and atmospheric neutrino problems are coming to an end
and we are entering an era of precision experiments. During the last
years, different results have given strong evidences of solar and
atmospheric neutrino oscillations~\cite{atm,solar}. Recently, the LMA
solution of the solar neutrino problem was confirmed by the KamLAND
reactor experiment~\cite{KamLAND} and also by more data from
SNO~\cite{TAUP}. The allowed regions for the solar and atmospheric
square mass differences and mixing parameters are, thus, getting very
constrained. We do not know, however, one very important question:
whether $\theta_{13}$, i.e., the $U_{e3}$ mixing, is different from
zero. This mixing is the door to the experimental measurement of
fundamental CP (or T) violation effects~\cite{CP}, the type of mass
hierarchy~\cite{mantle,core,mcwater} and controls the Earth matter
effect in supernova neutrino oscillations (see, e.g.,
Refs.~\cite{supernova}). Besides the experimental implications, the
smallness of $\theta_{13}$~\cite{CHOOZ,PaloVerde}, compared to the
other two mixing angles (in a three-neutrino mixing scheme), which are
relatively large~\cite{atm,solar,KamLAND}, is something not yet
explained from the theoretical point of view.    

The CHOOZ reactor experiment provides the more stringent bounds on the
value of $\theta_{13}$~\cite{CHOOZ}, although there are several
experiments consisting on conventional beams like K2K~\cite{K2K},
MINOS~\cite{MINOS} or CNGS experiments~\cite{CNGS}, which could
establish $\theta_{13} \neq 0$ or improve the present lower limit,
$\sin^2{}{2 \theta_{13}} < 0.10$. Even better limits are foreseen with
superbeams~\cite{superbeams} or neutrino factories~\cite{nufact}. 

When talking about controlled neutrino oscillation experiments, there 
are essentially two types of them: appearance and disappearance 
experiments. In an appearance experiment, a neutrino of a given flavor 
is produced. During the propagation, its flavor changes and it is 
detected via a pure charged current reaction. On the other hand, in a 
disappearance experiment a neutrino of a definite flavor is produced in
a controlled way and the depletion in the original flux after 
propagation is the signal for oscillation. The decrease in the original
flux is measured via charged current reactions which see the same flavor
as the one produced. However, for small mixings, the main signal in
the detector comes from neutrinos of the same flavor as the one
produced, so this means that there is less sensitivity to small
mixings for disappearance experiments. In addition, charged current
detection has a threshold energy for production, so that it is, in
general, impossible to use low energy neutrinos for appearance
experiments. 

It has been recently proposed~\cite{reacsup,reactor2det} the use of
reactor neutrinos to improve the sensitivity to $\theta_{13}$. In
order to do this, two detectors have to be used, a near detector and a
far detector. The latter at a distance of $\sim$ 1.7 km and the former
nearer so that no oscillations take place. In this way, systematic
errors can be reduced and a sensitivity down to $\sin^2{}{2
\theta_{13}} \simeq 0.01-0.02$ could be reached.  

Nuclear reactors produce low energy $\overline{\nu}_e$ and the basic
detection reaction is the inverse $\beta$-decay which has an energy
threshold of 1.806 MeV~\cite{reactorreview}. For these energies a
baseline of $\gtrsim$ 1 km is needed so that oscillations can take
place for the atmospheric square mass difference. This is, however, a
disappearance experiment and thus, less sensitive to small mixings,
which is the case. 

We would like to find an experiment capable of measuring very small
mixings. In order to accomplish this task, we will focus on the
following mixed charged and neutral current reaction:
$\overline{\nu}_e + e^- \rightarrow \overline{\nu}_e + e^-$. We will
make use of the fact that for another flavor, $\overline{\nu}_x + e^-
\rightarrow \overline{\nu}_x + e^-$ (with $x \neq e$) is a pure
neutral current reaction. Consequently, the cross sections for these
reactions are different. In principle, this fact could be used to
perform a neutrino oscillation experiment ($\overline{\nu}_e
\rightarrow \overline{\nu}_x$) which would be a mixture of appearance
and disappearance experiments. This mixture depends on the
neutrino energy and the electron recoil direction, so it could be
tuned by the choice of the appropiate kinematics. If oscillations take
place, the number of recoil electrons will be different from the case
of no oscillations. However, if both cross sections are similar, the
effect has a minor impact on the study of oscillations. Nevertheless,
it is known~\cite{zero} that the cross section for the scattering of
electron antineutrinos on electrons presents a destructive
interference and a dynamical zero for the kinematic configuration
corresponding to an incident antineutrino energy, $E_\nu =
\frac{m_e}{4 \sin^2{\theta_W}} \simeq m_e$, and maximum recoil energy
$T = T_{max} = \frac{2 E^{2}_{\nu}}{2 E_\nu + m_e} \simeq \frac{2
  m_e}{3}$ (forward electron). The point here is that this zero is not
present in $\overline{\nu}_x + e^- \rightarrow \overline{\nu}_x + e^-$
and this fact could make possible to perform an appearance-like
experiment. Indeed, if we were able to select only the events in a
window around the dynamical zero configuration, we would be detecting
almost only $\overline{\nu}_x$ and not $\overline{\nu}_e$ which would
be a sort of appearance-like experiment. We will take advantage of
these facts in order to study the possibilities of using this channel
to measure (or to get more restrictive bounds on) $U_{e3}$. For
typical antineutrino energies in a reactor, the inverse $\beta$-decay
reaction is the dominant one and the cross section for $\overline{\nu}
+ e \rightarrow \overline{\nu} + e$ is less than 1\% that of
$\overline{\nu}_e + p \rightarrow e^+ + n$. Nevertheless,
neutrino-elastic scattering has no energy threshold and the reactor
neutrino flux has a maximum at $\sim$ 0.5--1.0 MeV. Keeping all this
in mind, we will show that the near detector could be used to search
for oscillations in this channel, and not only to reduce systematic
errors in the far detector. Therefore, the combination of the
measurement in the near and in the far detectors might improve the
sensitivity to $\theta_{13}$. The main purpose of this paper is to
motivate this channel as a suitable way to look for oscillations in
the near detector.    

It is important to remark several additional facts which explain why
it is worthwhile to study more carefully this sort of appearance-like
experiment by means of the $\overline{\nu}_e-e^-$ reaction: 

\begin{description}

\item{i)} The dynamical zero is only present for $\bar{\nu}_{e}$, not
for $\nu_{e}$ or $\nu_{\mu}$ ($\bar{\nu}_{\mu}$), $\nu_{\tau}$
($\bar{\nu}_{\tau}$).

\item{ii)} The flavour $\bar{\nu}_{e}$ is precisely the one which is
produced copiously in nuclear reactors.

\item{iii)} The neutrino energy at which the zero appears is around the
peak of the antineutrino reactor spectrum~\cite{Vog-Eng89,reactorflux}. 

\item{iv)}  The dynamical zero is located at the maximum electron
recoil energy $T\simeq 2m_{e}/3$. This value is in the range of the
proposed experiments~\cite{BOREXINO} to detect recoil electrons.

\end{description}

The outline of the paper is the following. In section II, we present
the framework of neutrino oscillations. In section III we present the
basics of the far detector measurement. In section IV, we analyze the
optimal baseline for the near detector in order to be sensitive to
neutrino oscillations working as an appearance experiment. The
sensitivity to $\theta_{13}$, comparatively as what can be done just
with the far detector, is studied. Finally, in section V, we present our
conclusions.


\section{Reactor Neutrino Oscillations}

In the case of reactor neutrino experiments we are dealing with short
baselines and thus, when considering neutrino oscillations, we can
safely neglect matter effects. The form for the survival probability is
then given by  

\begin{widetext}
\begin{eqnarray}
\label{pee}
P_{\bar{\nu}_e \rightarrow \bar{\nu}_e} & = & 1 -
\cos^{4}{\theta_{13}} \sin^{2}{2 \theta_{12}} \,
\sin^{2}{\left(\frac{\Delta m^{2}_{21} L}{4 E}\right)} \nonumber\\ & & +
\sin^{2}{2 \theta_{13}} \left[ \cos^{2}{2 \theta_{12}} \,
\sin^{2}{\left(\frac{\Delta m^{2}_{31} L}{4 E}\right)} + 
\sin^{2}{\theta_{12}} \, 
\sin^{2}{\left(\frac{\Delta m^{2}_{32} L}{4 E}\right)} \right]
\end{eqnarray}
\end{widetext}

Considering $\sin^{2}{2 \theta_{13}}$ and 
$\sin^{2}{\left(\frac{\Delta m^{2}_{21} L}{4 E}\right)}$ small, to the 
first order in this approximation we can write Eq.~(\ref{pee}) as

\begin{widetext}
\begin{equation}
\label{peeapprox}
P_{\bar{\nu}_e \rightarrow \bar{\nu}_e} \simeq 1 - \left[ \sin^{2}{2
\theta_{13}} \, \sin^{2}{\left(\frac{\Delta m^{2}_{31} L}{4 E}\right)}
+ \sin^{2}{2 \theta_{12}} \,
\sin^{2}{\left(\frac{\Delta m^{2}_{21} L}{4 E}\right)} \right] 
\end{equation}
\end{widetext}

\noindent
If the high $\Delta m^{2}_{21}$ solution, with $\Delta m^{2}_{21} \sim
10^{-4}$ eV$^2$,  had turned out to be the right one, special care for
the second term in the bracket in Eq.~(\ref{peeapprox}) would have been
needed. In this case the determination of $\theta_{13}$ and
$\theta_{12}$ are coupled, and a joint analysis of reactor
antineutrino experiments with baseline of about 1 km and KamLAND would
be needed (see Ref.~\cite{th13KamLAND} for a study of the impact of
$\theta_{13} \ne 0$ on KamLAND data). The new SNO salt phase
data~\cite{TAUP}, however, strogly points towards the low $\Delta
m^{2}_{21}$ solution. 

From the simplycity of Eq.~(\ref{peeapprox}) it is easily seen that
correlations and degeneracies play a minor role in these type of
experiments. However, this means that there exist some limitations, as
it is the fact that there is no dependence on the atmospheric neutrino
mixing $\theta_{23}$, on the type of hierarchy (sign of $\Delta
m^{2}_{31}$) or on the CP violating phase.

Throughout the paper we will use the following values for the different
neutrino oscillation parameters~\cite{atm,TAUP}: 
 
\begin{eqnarray}
\Delta m^{2}_{21} = 7.1 \times 10^{-5}\mbox{eV}^{2} \hspace{5mm}; &
\Delta m^{2}_{31} = 3.0 \times 10^{-3}\mbox{eV}^{2} \hspace{5mm}; & 
\tan^{2}{\theta_{12}}= 0.41 
\end{eqnarray}

The bracket in Eq.~(\ref{peeapprox}), giving the appearance
probability, $P_{\bar{\nu}_e \rightarrow \bar{\nu}_x}$ ($x \ne e$),
shows its sensitivity to small values of the mixing angle
$\theta_{13}$, unlike the case of the disappearance channel.


\section{Far vs near detector}

We will first consider the basics of the far detector reaction and
the use of the elastic antineutrino-electron scattering in the
near detector.

In order to reach a good sensitivity to $\sin^{2}{2 \theta_{13}}$, the
detection of small spectral distortions in the positron event rates
due to antineutrino oscillations is important. This is only possible
by selecting an optimized baseline and by reducing systematic
uncertainties to the level of 1\%. These two points are crucial if we 
want to achieve an order of magnitude of improvement for the
$\sin^{2}{2 \theta_{13}}$ sensitivity. In the case of the far detector
the dominant detection reaction is the inverse $\beta$-decay 

\begin{equation}
\label{beta}
\overline{\nu}_e + p \longrightarrow e^+ + n  \hspace{2cm}
(E_{\nu})_{th} = \mbox{1.804 MeV}
\end{equation}

The selection of the proper baseline which gives the first oscillation
maximum for reactor antineutrinos, directly follows from the typical
energies of the inverse $\beta$-decay reaction, i.e., 3.5--4.0 MeV. As
we will see below, for $\Delta m^{2}_{31} = (2.5 - 3.0) \times
10^{-3}$ eV$^2$ the optimum baseline is $\sim$ 1.7 km. 

We assume a far detector technology like the CHOOZ or KamLAND
detectors and a typical integrated luminosity of $\mathcal{L} =$
8000 t $\cdot$ GW $\cdot$ yr. For concreteness, in the case of the
Kashiwazaki-Kariwa nuclear power plant whose maximum thermal power is
24.3 GW, and a 100-ton detector, an exposure-time of $\sim$ 3.3 years
would represent that luminosity.   

Reaction~(\ref{beta}) has a easily recognizable signal, the positron
anihilation with an electron, in delayed coincidence with the
$\gamma-$ray from the neutron capture. The energy of the positron is
given by

\begin{equation}
\label{energy}
E_{e^+} = E_{\overline{\nu}_e} - (M_n - M_p) +
O(E_{\overline{\nu}_e}/M_n) \simeq E_{\overline{\nu}_e} - 1.293 \mbox{
  MeV} 
\end{equation}

The visible energy in the detector is given by the sum of the positron
energy plus the mass of the anihilated electron, $E_{vis} = E_{e^+} +
0.511$ MeV. Therefore, a precise measurement of $E_{vis}$ corresponds
to a precise determination of the neutrino energy, $E_{\bar{\nu}_e}$.
Considering constant detector efficiency, $\epsilon$, the expected
number of events in the detector is given by 

\begin{equation}
\label{events}
N = \mathcal{N}_p \times \mathcal{T}_{exp} \times \epsilon \times
\frac{1}{4 \pi L^2} \times \int \frac{d\Phi}{dE_{\overline{\nu}_e}}
(E_{\overline{\nu}_e}) \cdot \sigma(E_{\overline{\nu}_e}) \cdot
P_{\overline{\nu}_e \rightarrow \overline{\nu}_e}
(E_{\overline{\nu}_e}) \cdot dE_{\overline{\nu}_e} 
\end{equation}

\noindent
where $\mathcal{N}_p$ is the number of protons in the detector,
$\mathcal{T}_{exp}$ is the exposure-time, $L$ is the reactor-detector
distance, $d\Phi/dE_{\overline{\nu}_e} (E_{\overline{\nu}_e})$ is the
initial reactor energy spectrum, $\sigma(E_{\overline{\nu}_e})$ is the
cross section for inverse $\beta$-decay and $P_{\overline{\nu}_e
  \rightarrow \overline{\nu}_e} (E_{\overline{\nu}_e})$ is the survival
$\overline{\nu}_e$ given by Eq.~(\ref{peeapprox}).  

The shape of the spectrum can be derived from a phenomenological
parameterization of the spectra from several of short baselines
experiments~\cite{reactorflux}  

\begin{equation}
\label{spectrum}
\frac{d\Phi}{dE_{\overline{\nu}_e}} = e^{a_0 + a_1
  E_{\overline{\nu}_e} + a_2 E_{\overline{\nu}_e}^2}
\end{equation} 
   
\noindent
where the values of the energy coefficients depend on the parent nuclear
isotope. This expression is a very good approximation for antineutrino
energies above 2 MeV. For lower energies, we have used a calculation
based on a summation of the allowed shape $\beta$ decays of all fission
fragments. The coefficients of Eq.~(\ref{spectrum}) and the calculated
spectra for lower energies are given in Ref.~\cite{Vog-Eng89}. In
addition, we assume a constant chemical composition for the reactor,
53.8\% of $^{235}$U, 32.8\% of $^{239}$Pu, 7.8\% of $^{238}$U and
5.6\% of $^{241}$Pu (see, e.g., Refs.~\cite{Bugey,th13KamLAND}). We will
also consider the thermal energy associated with the fissioning of
each of those nuclei as given in Ref.~\cite{Boe-Vog92}, that is 201.7
MeV for $^{235}$U, 205.0 MeV for $^{239}$Pu, 210.0 MeV for $^{238}$U
and 212.4 MeV for $^{241}$Pu.

\begin{figure}[t]
\begin{center}
\includegraphics[height=8cm]{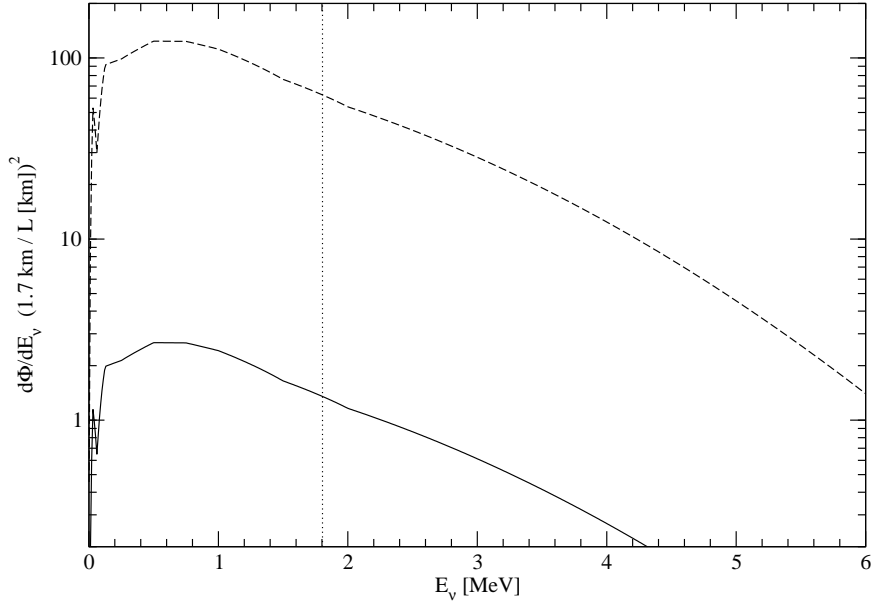}
\caption{\label{flux} Antineutrino nuclear reactor flux weighted by flux
  square-distance factor relative to the far detector at 1.7 km. Near
  detector at 0.25 km (dashed line) and far detector at 1.7 km (solid
  line) from the reactor.}
\label{flux}
\end{center}
\end{figure}  

In Fig.~\ref{flux}, we show the antineutrino nuclear reactor
flux for the near (dashed curve) and far (solid curve) detector
distances to the reactor, weighted by the flux square-distance factor in
each detector. As can be seen from the figure, the nuclear reactor
flux presents a maximum around $E_{\overline{\nu}} \simeq$  0.5--1.0
MeV, which it is roughly a factor of seven with respect to the
relevant energies in the far detector, $E_{\overline{\nu}}$ = 3.5--4.0
MeV. Thus, we have taken 1.7 km and 0.25 km ($\simeq$ 1.7 km/7) for
the far and near detector-reactor distances, respectively. The inverse
$\beta$-decay reaction is only sensitive to antineutrino energies
higher than the threshold one, 1.806 MeV (limited by the dotted line),
while the antineutrino-electron elastic reaction is so for the entire
spectrum. In addition, as can be seen from Fig.~\ref{flux}, within the
region of the maximum, the nuclear reactor flux is a few times larger
than for the relevant energies detected by the inverse $\beta$-decay
reaction in the far detector. All in all, due to being closer to the
nuclear reactor and working in a higher-flux region, the spectrum
around the maximum in the near detector is a factor $\sim$ 100 times
larger than the part of it sensitive to the inverse $\beta$-decay in
the far detector. This can be understood by comparing the dashed and
solid lines in Fig.~\ref{flux}.  

On the other hand, there are different calculations of the cross
section for the inverse $\beta$-decay which take into account
different approximations valid for different
regimes~\cite{crosssection}. To the lowest order, this cross section
is given by   

\begin{equation}
\label{cross}
\sigma(E_{e^+}) = \frac{2 \pi^2}{m^{5}_{e} \, f \, \tau_n} \, 
p_{e^+} E_{e^+} 
\end{equation}

\noindent
where $f$ is the phase space factor for the free neutron decay and
$\tau_n$ is the lifetime of a free neutron. 

Although the cross section for $\overline{\nu} + e \rightarrow
\overline{\nu} + e$ is about 1\% that of $\overline{\nu}_e + p
\rightarrow e^+ + n$, the flux gain, discussed before, due to the use
of the near detector around the maximum of the spectrum compensates
this factor. Therefore, we expect, roughly, a similar number of events
in the far detector using the inverse $\beta$-decay reaction and in
the near detector using the antineutrino-electron elastic
scattering\footnote{Assuming similar masses for both detectors.}.  

Thus, as the mixing $\theta_{13}$ is small and the far detector
performs a disappearance experiment, it is very important to reduce
systematic uncertainties. The near detector will help in this task
using the same reaction as the far detector, but it will also be
useful to perform neutrino oscillation studies by itself using
antineutrino-electron elastic scattering.


\section{Appearance-Like Experiment}

Many of the systematic uncertainties, due to poor knowledge of the
neutrino flux, number of protons and detection efficiency cancel out if
besides a far detector, a near detector is used and measurements
in both detectors are compared. It has been recently
shown~\cite{reacsup,reactor2det} that the use of a near detector at
$\sim$ 0.2 km makes possible the determination of
$\sin^{2}{2\theta_{13}}$ down to 0.01--0.02. It has also been shown that
reactor measurements can play a role complementary to long baseline
experiments, helping to resolve parameter degeneracies.  

As we have already argued above, we will show that not only is the
near detector useful to lower the systematic uncertainties, but also
to perform neutrino oscillation measurements complementary to those in
the far detector, by using antineutrino-electron elastic scattering for
energies around the maximum of the reactor antineutrino spectrum,
which, combined with the smaller baseline, implies a flux gain of
$\sim$ 100 with respect the far detector measurements. Thus, although
the antineutrino-electron elastic cross section is a factor $\sim$ 100
smaller than in the case of inverse $\beta$-decay, working on the
maximum of the reactor spectrum, allows us to use this reaction for
neutrino oscillation studies in the near detector.     

The main purpose of using the antineutrino-electron elastic scattering
as the detection reaction is to simulate an appearance experiment. In
order to achieve this, only that part of the recoil electron spectrum
around the dynamical zero~\cite{novel} must be considered. For this
channel, the number of events is given by 

\begin{widetext}
\begin{equation}
\label{N}
N = \mathcal{N}_p \times \mathcal{T}_{exp} \times \epsilon \times
\frac{1}{4 \pi L^2} \times \int \frac{d \sigma^{\overline{\nu}}}{dT}
(E_{\overline{\nu}_e},T) \cdot  \frac{d \phi^o}{dE_{\overline{\nu}_e}}
(E_{\overline{\nu}_e}) \cdot dE_{\overline{\nu}_e} \, dT   
\end{equation}  
\end{widetext} 

\noindent
where $\frac{d \sigma^{\overline{\nu}}}{dT} (E_{\overline{\nu}_e},T)$
is the sum of all the cross sections convoluted with the oscillation 
probabilities\footnote{We are taking the differential cross sections
  for  $\overline{\nu}_\mu$ and $\overline{\nu}_\tau$
  ($\overline{\nu}_x$) as equal, not considering radiative corrections.}

\begin{widetext}
\begin{equation}
\label{dsdT}
\frac{d \sigma^{\overline{\nu}}}{dT} (E_{\overline{\nu}_e},T) = 
P_{\overline{\nu}_e \rightarrow \overline{\nu}_e}
(E_{\overline{\nu}_e}) \,  \frac{d \sigma^{\overline{\nu}_e}}{dT}
(E_{\overline{\nu}_e},T) + P_{\overline{\nu}_e \rightarrow
  \overline{\nu}_x} (E_{\overline{\nu}_e}) \, \frac{d
  \sigma^{\overline{\nu}_x}}{dT} (E_{\overline{\nu}_e},T) 
\end{equation}
\end{widetext}

The first term in Eq.~(\ref{dsdT}), the dissapearance term, is the one
measured in the far detector, but it cannot be substracted out because
the energies of interest in the near detector are much lower for the
elastic reaction. Around the dynamical zero, $d\sigma^{\bar{\nu}_e}/dT
= 0$, and Eq.~(\ref{dsdT}) shows that around this point, this reaction
simulates an appearance-like experiment. 

Using the fact that the probability of $\overline{\nu}_e$ going to an
antineutrino of any flavor must be equal to one, we can rewrite
Eq.~(\ref{dsdT}) as   

\begin{widetext}
\begin{equation}
\label{dsdTx}
\frac{d \sigma^{\overline{\nu}}}{dT} (E_{\overline{\nu}_e},T) =  
\frac{d \sigma^{\overline{\nu}_e}}{dT} (E_{\overline{\nu}_e},T) + \left(
\frac{d \sigma^{\overline{\nu}_x}}{dT} (E_{\overline{\nu}_e},T) - 
\frac{d \sigma^{\overline{\nu}_e}}{dT} (E_{\overline{\nu}_e},T)
\right) \, P_{\overline{\nu}_e \rightarrow \overline{\nu}_x}  
\end{equation}  
\end{widetext}

The antineutrino-electron elastic scattering cross sections are given
by~\cite{Vog-Eng89}

\begin{equation}
\label{ds}
\frac{d\sigma^{\overline{\nu}_i}}{dT} (E_{\overline{\nu}_i},T)=
\frac{2 \, G_F \, m_e}{\pi} \, \left[(g^{i}_{R})^2 + (g^{i}_{L})^2 \,
  \left( 1 - \frac{T}{E_{\overline{\nu}_i}} \right)^2 - g^{i}_{L} \,
  g^{i}_{R} \, \frac{m_e \, T}{E^{2}_{\overline{\nu}_i}} \right] 
\end{equation}

\noindent
where $G_F$ is the Fermi coupling constant, $T$ the recoil kinetic 
energy of the electron and $E_{\overline{\nu}_i}$ the antineutrino
incident energy. For neutrinos one has to make the change 
$g^{i}_{L} \leftrightarrow g^{i}_{R}$. In terms of the weak mixing angle
$\theta_W$, the chiral couplings $g^{i}_{L}$ and $g^{i}_{R}$ can be 
written for each neutrino flavor as

\begin{widetext}
\begin{eqnarray}
\label{g}
g^{e}_{L} = \frac{1}{2} + \sin^{2}{\theta_W} & \hspace{1cm}g^{e}_{R} = 
\sin^{2}{\theta_W} \nonumber\\ & \\
g^{\mu, \tau}_{L} = - \frac{1}{2} + \sin^{2}{\theta_W} & \hspace{1cm}
g^{\mu, \tau}_{R} = \sin^{2}{\theta_W} \nonumber
\end{eqnarray} 
\end{widetext}

From Eq.~(\ref{ds}) it is evident that if $g_{L}^i g_{R}^i>0$ there is
a chance for the cross section to cancel in the physical region. From
Eq.~(\ref{g}) we see that this zero is only possible in the
$\overline{\nu}_{e} e^- \rightarrow \overline{\nu}_{e} e^-$ channel
and, in fact, it takes place for the kinematical configuration
$E_{\nu}=m_{e}/(4 \sin^2\theta_{W})$ and maximal $T$. Neither
 $d\sigma^{\overline{\nu}_{\mu}}/dT$ nor
$d\sigma^{\overline{\nu}_{\tau}}/dT$  present a dynamical zero since
$g_{L}^{\mu ,\tau}g_{R}^{\mu ,\tau}<0$. We will take advantage of this
fact.   

\begin{figure}[t]
\begin{center}
\includegraphics[height=12cm]{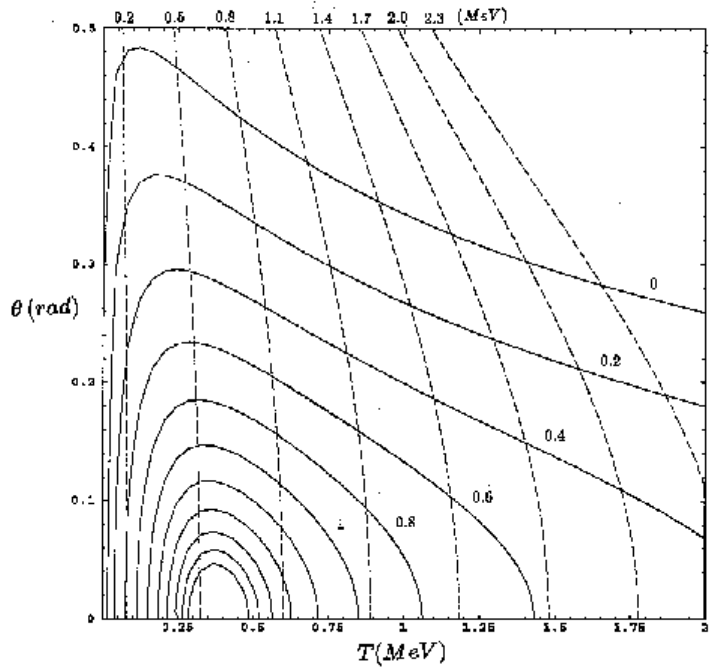}
\caption{\label{d} Curves of constant values of $d \equiv
	\log{\left[\frac{ \sigma^{\overline{\nu}_\mu}}{dT}/\frac{d
	\sigma^{\overline{\nu}_e}}{dT}\right]}$ (solid lines) in the
  plane ($\theta, T$). Curves of constant antineutrino energy are also
	plotted (dashed lines).}  
\end{center}
\end{figure}

In Fig.~\ref{d} we present the curves of constant values of $d \equiv
\log{\left[\frac{d \sigma^{\overline{\nu}_\mu}}{dT}/ \frac{d
      \sigma^{\overline{\nu}_e}}{dT}\right]}$ (solid lines) in the
plane ($\theta, T$) where the different regions where the appearance
channel starts to be important\footnote{The relation between $\theta$
  and the other two kinematic variables, $E_{\overline{\nu}}$ and $T$, 
  is given by $\cos{\theta} = \frac{E_{\overline{\nu}} +
    m_e}{E_{\overline{\nu}}} \sqrt{\frac{T}{T + 2 m_e}}$.}, that is when
$\frac{d \sigma^{\overline{\nu}_\mu}}{dT} > \frac{d
  \sigma^{\overline{\nu}_e}}{dT}$, can be clearly seen. Curves of
constant antineutrino energy are also shown (dashed lines). 

\begin{figure}[t]
\begin{center}
\includegraphics[height=8cm]{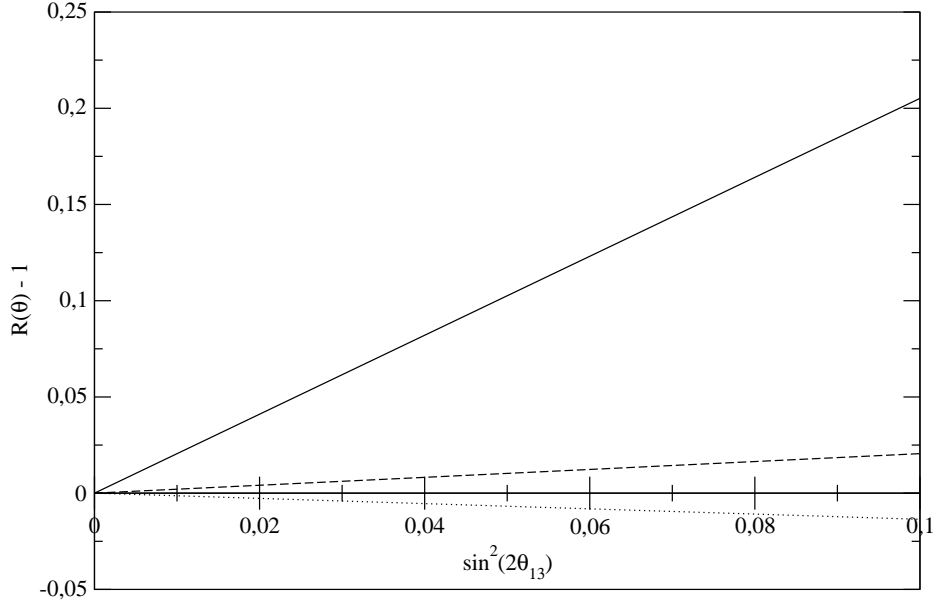}
\caption{\label{R-1} $R(\theta) - 1$ as a function of $\sin^{2}{2
  \theta_{13}}$, for a given T-interval, T $\in$ [0.25, 0.80] MeV, a
  reactor-detector distance of 0.25 km, and for different values of
  $\theta$ = 0.3 rad (solid line), 0.5 rad (dashed line) and
  $\theta_{max}$ = 1.11 rad (dotted line).}  
\end{center}
\end{figure}

Let us now consider the following observable: 

\begin{equation}
\label{R}
R(\theta) = \frac{N (\theta)}{N_{U_{e3}=0} (\theta)}
\end{equation}

\noindent
where $N (\theta)$ is the number of events for electron recoil angles 
smaller than $\theta$ in the case of oscillations and $N_{U_{e3}=0}
(\theta)$ is the corresponding prediction for $U_{e3}=0$. Close to the
configuration of the dynamical zero (small $\theta$ and the T-interval
around $T \simeq 2 m_e / 3$) $R > 1$ (appearance-like experiment),
while if we consider a bigger sample $R < 1$ (disappearance-like
experiment). This can be clearly seen by plotting $R(\theta) - 1$ for
different values of $\theta$ = 0.3 (solid line), 0.5 (dashed line) and
1.11 $\equiv \theta_{max}$ (dotted line) rad, as a function of
$\sin^{2}{2 \theta_{13}}$, Fig.~\ref{R-1}, for an electron recoil
energy interval T $\in$ [0.25, 0.80] MeV and for a reactor-detector
baseline of L = 0.25 km.  

As can be seen from Fig.~\ref{R-1}, the region around the dynamical zero
has a much better sensitivity to $U_{e3}$ than in the case of making
no angular selection. As expected, this is due to the fact that in
that case, an appearance-like experiment is simulated, which is much
more sensitive to small mixings than a disappearance-like one. Of
course, when narrowing the angular detection window, the statistics is
smaller. The immediate question one should wonder looking at
Fig.~\ref{R-1} is whether this gain in sensitivity to $\theta_{13}$
when considering small regions is large enough to compensate this
decrease of statistics. It is important to notice that this high slope
can be due to the ratio between two small quantities, being the
denominator close to zero (it nearly vanishes over the dynamical
zero). From Fig.~\ref{R-1}, it can also be seen that there is an
intermediate region where the effect of $\overline{\nu}_e$ and
$\overline{\nu}_x$ interfere destructively and we have no sensitivity
at all to $\theta_{13}$, even having much more statistics and
independently of the value of $\theta_{13}$. This is clear from
Eq.~(\ref{dsdTx}), for the term that depend on $\theta_{13}$ will be
suppressed when the effect of $\frac{d \sigma^{\overline{\nu}_e}}{dT}
(E_{\overline{\nu}_e},T)$ and $\frac{d \sigma^{\overline{\nu}_x}}{dT}
(E_{\overline{\nu}_e},T)$ compensate each other. Even if there are
oscillations, at that configuration, the number of events is given
just by $d \sigma^{\overline{\nu}_e} / d T$, and hence not being
sensitive to oscillations. As seen from Eq.~(\ref{dsdTx}), the
kinematics of that cancellation is given by the only condition that
the charged current amplitude is twice the neutral current one,
whereas the dynamical zero of the $\overline{\nu}_e$ reaction shows up
when the two interfering amplitudes are equal. 

\begin{figure}[t]
\begin{center}
\includegraphics[height=8cm]{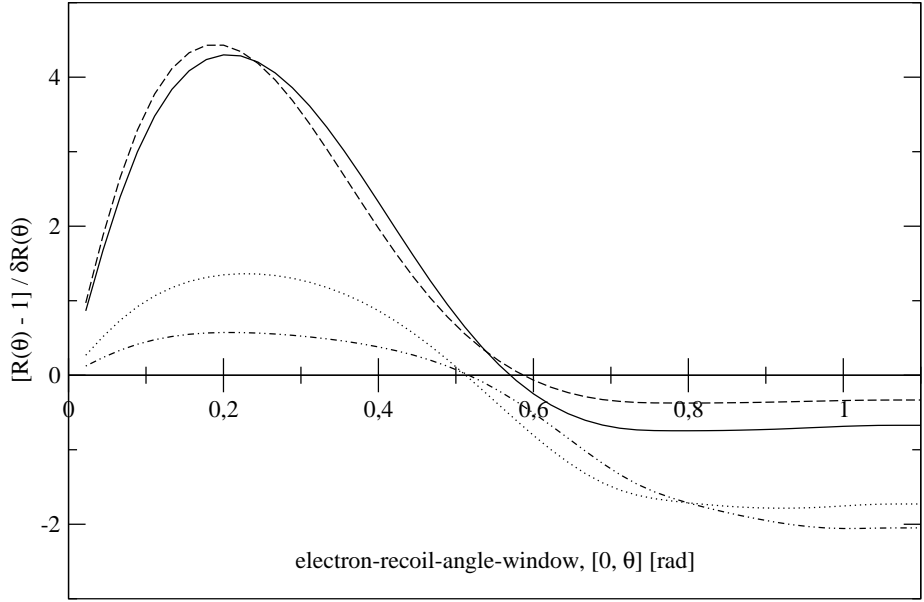}
\caption{$(R(\theta) - 1)/\delta R(\theta)$ as a function of the
  electron recoil angle window (from 0 rad to $\theta$ rad) for T $\in$
  [0.25, 0.80] MeV, $\sin^{2}{2\theta_{13}}$ = 0.04 and a
  reactor-detector baseline of 0.17 km (dashed line), 0.25 km (solid
  line), 0.50 km (dotted line) and 0.75 km (dot-dashed line).}  
\label{sensi}
\end{center}
\end{figure}

In order to estimate the bounds one could extract by measuring
$R(\theta)$, we will assume that using the near detector to reduce
systematics with the inverse $\beta$-decay reaction, lowers the
uncertainty on the normalization of the reactor flux to $\sigma_{sys}$
= 0.8\%~\cite{reactor2det}. Then, for a semi-quantitative analysis, we
will assume only statistical errors along with this systematic one
associated to the normalization of the antineutrino spectrum. In
Fig.~\ref{sensi}, $(R(\theta) - 1)/\delta R(\theta)$ is shown as a
function of the electron-recoil-angle-window, $\theta$-window, within
the T-interval, T $\in$ [0.25, 0.80] MeV, for different reactor-detector
distances and $\sin^{2}{2\theta_{13}}$ = 0.04. From Fig.~\ref{sensi},
it is evident that if the entire $\theta$-window is considered 
(dissapearance regime), the sensitivity to small $U_{e3}$ decreases as
the reactor-detector distance decreases. On the contrary, this is the
opposite to what happens within a $\theta$-window around the dynamical
zero (appearance regime), the sensitivity increases as the
reactor-detector distance decreases up to $\sim$ 0.15--0.25 km.   

These opposite behaviors can be understood by the fact that in the
disappearance regime larger antineutrino energies play a role, and
thus larger distances keep the oscillatory factor in the probability
around its maximum. On the other hand, for antineutrino energies
around the dynamical zero, this oscillatory maximum is reached at a
reactor-detector distance of $\sim$ 0.25 km. 

\begin{figure}[t]
\begin{center}
\includegraphics[height=8cm]{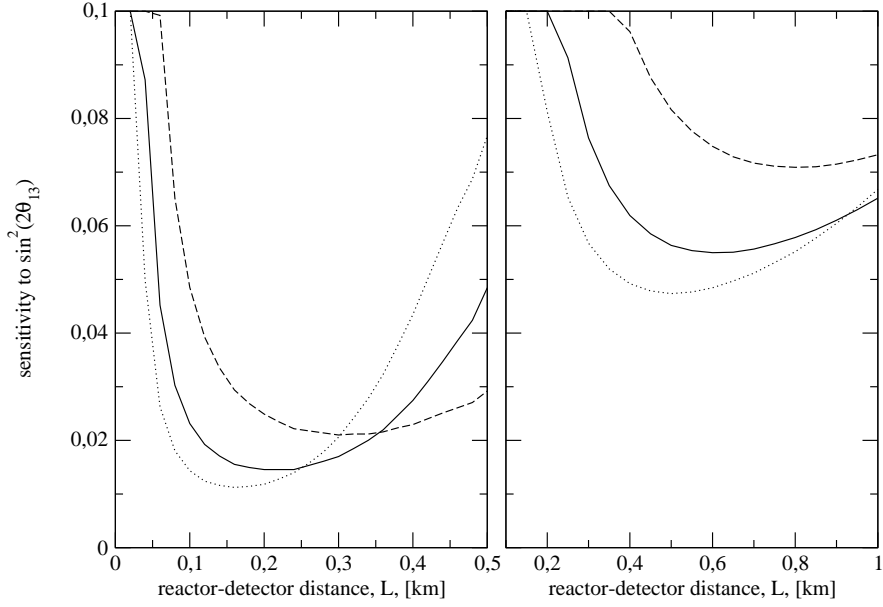}
\caption{Sensitivity to $\sin^{2}{2\theta_{13}}$ as a function of the
  reactor-detector baseline, L (in km), for different values of
  $\Delta m^{2}_{31}$ = 2 $\times$ 10$^{-3}$ eV$^{2}$ (dashed line), 3
  $\times$ 10$^{-3}$ eV$^{2}$ (solid line) and 4 $\times$ 10$^{-3}$
  eV$^{2}$ (dotted line), at the 90 \% confidence level. The detection
  window is T $\in$ [0.25, 0.80] MeV and: $\theta$ $\in$ [0, 0.25] rad
  (left plot) and for all $\theta$ (right plot).}  
\label{distance}
\end{center}
\end{figure} 

Comparatively, the best configuration for the appearance regime
($\sim$ 0.25 km) gives a factor of two better in $(R(\theta) -
1)/\delta R(\theta)$ than the best configuration
for the disappearance regime ($\sim$ 0.75 km). For L = 0.25
km, $\sin^{2}{2\theta_{13}}$ = 0.04 would be resolved with a 4$\sigma$
confidence level in the appearance regime. For this configuration, a
sensitivity down to $\sin^{2}{2\theta_{13}}$ = 0.015 could be reached
at 90\% confidence level, which is comparable to the sensitivity that
can be reached using the inverse $\beta$-decay reaction in the far
detector, $\sin^{2}{2\theta_{13}} \sim$ 0.01. This is shown in the
left plot of Fig.~\ref{distance} where the sensitivity to
$\sin^{2}{2\theta_{13}}$ (largest value of $\sin^{2}{2\theta_{13}}$
which fits the value $\sin^{2}{2\theta_{13}}$ = 0 at the chosen
confidence level) is depicted as a function of the reactor-detector
baseline, L (in km), for different values of $\Delta m^{2}_{31}$ = 2
$\times$ 10$^{-3}$ eV$^{2}$ (dashed line), 3 $\times$ 10$^{-3}$
eV$^{2}$ (solid line) and 4 $\times$ 10$^{-3}$ eV$^{2}$ (dotted line),
at the 90 \% confidence level. The detection window selected is
$\theta$ $\in$ [0, 0.25] rad and T $\in$ [0.25, 0.80] MeV. Thus, in
what the dependence with $\Delta m^{2}_{31}$ is concerned, there is a
worse (better) sensitivity as it decreases (increases), within the
allowed experimental range. For smaller values of $\Delta m^{2}_{31}$
the sensitivity becomes slightly worse as the reactor-detector baseline
becomes shorter (within L = 0.15--0.25 km), while for larger values of
$\Delta m^{2}_{31}$, the best sensitivity is obtained at shorter
baselines. The right plot in Fig.~\ref{distance}, analogous to the
left one but with no $\theta$-window selected, i.e., counting all the
events, shows in another way that the dissapearance channel is less
sensitive for the allowed range of neutrino oscillation
parameters. Thus, the knowledge of the kinematic region around the
dynamical zero is of crucial importance in order to achieve a
comparable sensitivity to the one in the far detector.   
 
\begin{figure}[t]
\begin{center}
\includegraphics[height=8cm]{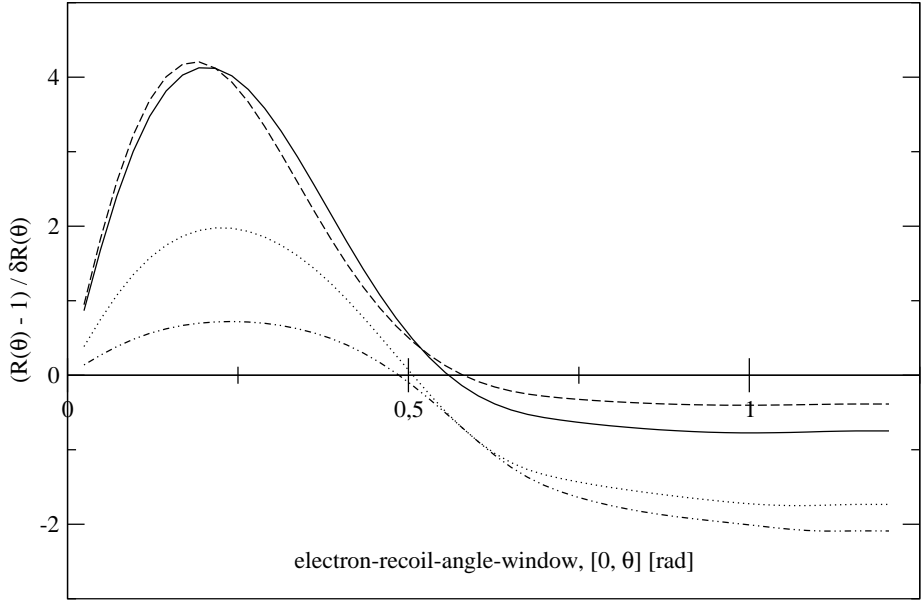}
\caption{$(R(\theta) - 1)/\delta R(\theta)$ as a function of the
  electron-recoil-angle-window (from 0 rad to $\theta$ rad) for T
  $\in$ [0.15, 1.00] MeV, $\sin^{2}{2\theta_{13}}$ = 0.04 and a 
  reactor-detector baseline of 0.17 km (dashed line), 0.25 km (solid
  line), 0.50 km (dotted line) and 0.75 km (dot-dashed line).}  
\label{sensi2}
\end{center}
\end{figure} 

We can also study how the widening of the T-interval, keeping an
angular-window fixed, affects the sensitivity. This is shown in
Fig.~\ref{sensi2}, which is analogous to Fig.~\ref{sensi} but for T
$\in$ [0.15, 1.00] MeV. As we can see from the plot, considering
slightly wider T-intervals does not affect significantly the
sensitivity to $U_{e3}$ while having more events. If we keep on
widening the T-interval the sensitivity to $U_{e3}$ within the
appearance regime will decrease in a significant way for L $\sim$
0.15--0.25 km and will increase for L $\gtrsim$ 0.5 km. This occurs
because of the contribution of higher energies and the displacement of
the oscillation maximum to longer baselines. From a certain baseline,
L $\sim$ 0.75 km, and up, the lack of events decreases the
sensitivity. Thus, for a given $\Delta m^{2}_{31}$, there should be a
compromise between narrowing the detection window in order to consider
a region dominated by the dynamical zero (locating the detector at
$\sim$ 0.15-0.25 km) with a relative small number of events, and
opening up this window in order to have a larger number of events, and
then considering higher antineutrino energies having to move the
detector to longer baselines, and consequently losing flux. We have
found that the $\theta$-window, up to $0.2-0.3$ rad, is demanded,
whereas the recoil electron energy interval can be moderately extended
at the expense of increasing the baseline. An appropiate choice
appears for T $\in$[0.25, 0.80] MeV and L = 0.25 km.


\section{Conclusions}

Recent analyses have shown~\cite{reacsup,reactor2det} the interest of
using two detectors in reactor neutrino oscillation experiments in
order to reduce systematic errors and reach a sensitivity to the
$U_{e3}$ mixing comparable to the first-generation superbeams. Besides
this strategy, we propose in this paper the use of the near detector
to perform an appearance-like experiment by means of sitting around
the dynamical zero in the $\overline{\nu}_{e}-e$ elastic scattering
cross section~\cite{zero}. Although the cross section for
$\overline{\nu}_e + e^- \rightarrow \overline{\nu}_e + e^-$ is about
1\% that of $\overline{\nu}_e + p \rightarrow e^+ + n$, the flux gain
at smaller energies (around $E_{\overline{\nu}_e} = 0.5$ MeV) and the
corresponding shorter baseline of the near detector compensate this
factor. 

For a configuration with the near detector at $\sim$ 0.25 km and a
window in the electron recoil angle for $\overline{\nu}_e + e^-
\rightarrow \overline{\nu}_e + e^-$ from 0 to $\sim$ 0.25 $rad$ (for
electron recoil kinetic energies up to $\sim$ 1 MeV), we find a
sensitivity down to $\sin^{2}{2 \theta_{13}}$ which is comparable to
the one that can be reached using the inverse $\beta$-decay reaction
in the far detector at 1.7 km. In those windows for $\overline{\nu}_e
+ e^- \rightarrow \overline{\nu}_e + e^-$, the cross section for
$\overline{\nu}_x$ ($x \ne e$) is larger than that for
$\overline{\nu}_e$ as can be seen in Fig.~\ref{d}.

\section*{ACKNOWLEDGMENTS}

We thank Massimo Passera and Thomas Schwetz for useful
discussions. This work is supported by the Spanish Grant FPA2002-00612
of the MCT. SPR has also been supported by the Spanish MECD for a FPU
fellowship and by NASA Grant ATP02-0000-0151.

\end{document}